\documentclass[11pt]{article}

\usepackage{graphicx}

\begin{document}

\centerline{\LARGE \bf Spectral Statistics of the 
Two-Body Random Ensemble Revisited}

\vspace*{1cm}

\centerline{\bf \Large J.~Flores$^{1}$, 
M.~Horoi$^{2}$, M.~M\"uller$^{3}$ and 
T.H.~Seligman$^{1}$}
\vspace*{.3cm}

{\it
\begin{center}
$^1$ Centro de Ciencias 
F\'{\i}sicas, UNAM, Campus Morelos, Cuernavaca, Mexico,
62251, A.P. 48-3\\

$^2$ Physics Department, Central Michigan University, Mount Pleasant,
MI 48859, USA\\

$^3$ Facultad de Ciencias, 
UAEM, C.P. 62210, Cuernavaca, Morelos, Mexico\\
\end{center}}

\vspace*{.5cm}

\vspace*{1.5cm}

\begin{abstract}
                
Using longer spectra 
we re-analyze spectral properties of the two-body random ensemble
studied thirty years ago. At the center of the spectra 
the old results are largely confirmed, 
and we show that the non-ergodicity is
essentially
due to the variance of the lowest moments of the spectra.
The longer spectra
allow to test and reach the limits of validity of French's correction
for the number variance.
At the edge of the spectra we discuss the problems 
of unfolding in more detail. With a Gaussian unfolding of each spectrum
the nearest neighbour spacing distribution between ground state and first exited
state is shown to be stable. Using such an unfolding the distribution 
tends toward a semi-Poisson distribution for longer spectra.
For comparison with the nuclear table ensemble we could use such unfolding
obtaining similar results as in the early papers, but an ensemble 
with realistic splitting gives reasonable
results if we just normalize the spacings in accordance with the procedure used
for the data.

\end{abstract}
\vspace{1cm}

\section{Introduction}

The Gaussian Orthogonal Ensemble (GOE) was originally introduced by Wigner 
\cite{Wi51} in 1951 into physics in order to describe statistics of 
isolated, high-lying nuclear levels. A detailed
analysis of each state seemed neither possible nor desirable.
Statistical properties, on the other hand, were needed for nuclear technology 
and were well described by the GOE. About a quarter of a century ago there was 
considerable interest in the so called embedded ensembles and in particular 
in the two-body random Hamiltonian ensemble (TBRE)
\cite{Fre70,1BohFlo71,Ger72,2BohFlo71},
because it was realized that the GOE represented an $n$-body interaction in a
mean
field basis, while it is generally assumed that nuclei can be fairly well 
described by an effective two-body interaction in this basis. 
Thus $n$-body Hamiltonians were constructed from two-body GOE's assuming
degenerate 
single-particle states. The spectral statistics of these two-body random
ensembles 
were analyzed. The main findings were:
\begin{enumerate}
\item
The level density of the TBRE is Gaussian \cite{Fre70,1BohFlo71,Ger72},
rather than semi-circular.

\item
The TBRE is neither stationary \cite{2BohFlo71} nor ergodic
\cite{Fre73,Brodetal}.

\item Unfolding the TBRE individually GOE statistics is recovered
\cite{Fre73,Brodetal}.

\item
The deviation with respect to GOE 
is due to the variance of average spacings over
the ensemble \cite{Fre73}.

\item
Fluctuations at the edge of the spectrum of a TBRE are very large 
\cite{2BohFlo71} compared to those of a GOE and can be roughly described by a
Brody
distribution \cite{Brod73} 
of nearest-neighbour spacings with a parameter $\omega = 0.85$
\cite{BroCoFloMe}.

\end{enumerate}

These results, with 
the exception of the one for the level density, are exclusively numerical.
They were obtained at a time of very limited computer facilities. As a
consequence,
only few particles were treated and the dimension of the TBRE matrices was
small.

There has been recently a flurry of interest in the TBRE regarding properties
of wave
 functions and spectra \cite{FlamGriIz} - \cite{Ber2}. 
In this context the old work was ignored to a large extent.
Yet it seems worthwhile to take it into account, because it is not clear what
effects 
non-ergodicity could have in the new context.
The numerical results can be improved considerably upon and by 
consequence the old results can be tested.
Furthermore it seems desirable to go beyond the TBRE to what we may call a 
realistic two-body random ensemble (RTBRE) by lifting the degeneracy condition 
on the single-particle spectrum.
Applications in nuclear physics and in other fields certainly require this.

In what follows we will first fix notation defining the TBRE properly and
proceed
in section 3 to analyze properties in the center of the spectrum. In the fourth
section we consider properties at the edge of the spectrum and 
we discuss the RTBRE which seems to differ mainly there.
Conclusions will be given in the last section.

\section{Definition of the TBRE}

The TBRE is defined for a fermionic system of $n$ spin $1/2$ particles in three
dimensions. A set of degenerate single-particle states with well defined
angular
momentum and other quantum numbers is used to construct $n$-particle states,
that belong to a good total angular momentum $J$, and if we think of a nucleus,
also a good isospin $T$. 
Due to the Pauli principle and the corresponding blocking of states,
the lowest single-particle states may not enter the picture because we assume
them to be filled and inert. The effective two-body interaction usually used 
in this context was replaced by a 
Gaussian distributed two-body interaction matrix 
whose strength is irrelevant as no energy scale is 
established by degenerate single-particle states.

For our numerical studies we choose a strength parameter
$\lambda$ such that  
the value $\lambda = 1$ corresponds to a typical interaction strength in the
nuclear
$2s - 1d$ shell on which we shall concentrate. 
This parameter 
will be important for the RTBRE, where realistic mean field parameters yield
the relevant
single-particle energies and thus provide an energy scale.
The calculations reported here were mainly carried out for eight particles in
the 
$2s - 1d$ shell.
Information about the quantum numbers, dimension of the spectra
and the number of matrices of each ensemble used 
can be found in Table
\ref{tabQN}.
The corresponding shell model calculations were carried out using the code 
OXBASH \cite{prog}.
As far as spectral statistics are concerned we shall mainly consider the 
nearest-neighbour spacing distribution $P(s)$ and the number variance
$\Sigma^2(L)$
and occasionally the skewness and the excess; for definitions we refer to a
standard book
such as Mehta \cite{Meht}.
\begin{table}
\caption{Quantum numbers, dimension of the matrices and the number of
members for the ensembles analyzed in this paper}
\label{tabQN}
\begin{center}
\begin{tabular}{|c||c|c|c|}
\hline
      & $(J,T)$ & dimension & number of matrices \\
\hline
\hline
 TBRE & $(0,0)$ & 325 & 500  \\
\hline
 TBRE & $(0,2)$ & 287 & 500  \\
\hline
 TBRE & $(2,0)$ & 1206 & 708 \\
\hline
RTBRE & $(0,0)$ & 325 & 500  \\
\hline
\end{tabular}
\end{center}
\end{table}

\section{Statistics in the center of the TBRE and non-ergodicity}

\begin{figure}
\begin{center}
\includegraphics[scale=0.6]{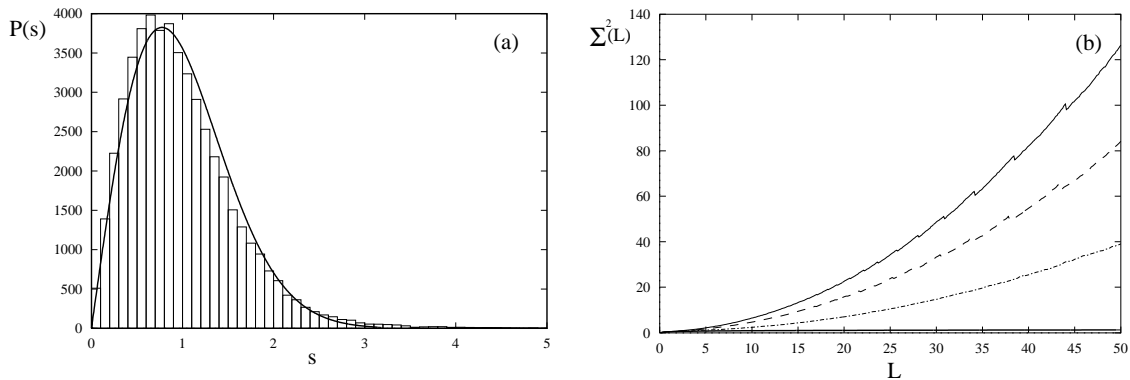}
\caption{(a) Ensemble averaged nearest-neighbour spacing 
distribution of the TBRE 
$(J=0, T=0)$ 
at the center of the spectrum. The full line corresponds to the GOE.
(b) Number variance $\Sigma^2(L)$ 
for the same case. The thin solid line corresponds
to the ensemble average, the
dashed line to the ensemble average after normalizing the 
spectrum widths and the dashed-dotted curve to the
ensemble average after re-centering the spectra. 
The GOE values are indicated by the thick solid line.}
\end{center}
\end{figure}
Before we proceed to analyze statistics of spectral fluctuations we have
to unfold the spectrum \cite{Meht} in order to have average spacing $D=1$
throughout the spectrum. For this purpose we could use the analytic result
that we have a Gaussian density for the TBRE \cite{Ger72}.
For practical purposes 
it turns out to be convenient simply 
to make a polynomial unfolding, using a best
fit
to the cumulative level density of the ensemble. For the central 
region of the spectrum the type of unfolding used is 
indeed not relevant.

We thus superpose all spectra and the integrated level density 
(staircase) is fitted with a polynomial of degree 15. 
Each spectrum is unfolded
with this polynomial for an energy interval at the center of the of the
spectra which contains on the average 115 states
and calculated 
$P(s)$ and $\Sigma^2(L)$. 
The results are shown in Figs. 1a and 1b. The short-range
fluctuations 
deviate significantly from the GOE and the long-range correlation show huge
deviations from this ensemble.
This becomes rather obvious if we look at the distribution of the average
energy and width of each spectrum individually.
In Table \ref{tab1} we show the values of the standard deviation of
the distribution of the average energy $\sigma_c$ and of the
widths distribution of the spectra $\sigma_v$ and compare
them with the widths $\sigma_{tot}$ of the 
energy distribution of the entire ensemble.
The scattering of the centers  and the widths 
of the TBRE and the RTBRE
in units of the total widths 
is about 30 times larger than the one of the GOE.
Both quantities fluctuate widely while for the GOE the 
distributions are extremely
narrow. 

\begin{table}
\caption{Standard deviation of the center distribution
$\sigma_c$, the distribution of the widths of the spectra 
$\sigma_v$ and of the total level density of the ensemble $\sigma_{tot}$
for three different TBRE's and the GOE}
\label{tab1}
\begin{center}
\begin{tabular}{|c||c|c|c|c|c|}
\hline
 & $\sigma_c$ & $\sigma_v$ & $\sigma_{tot}$ & $\sigma_c/\sigma_{tot}$ &
 $\sigma_v/\sigma_{tot}$ \\
\hline
\hline
 TBRE $J=0, T=0$ & $20.1$ & $4.5$ & $39.7$ & $0.51$ & $0.11$ \\
\hline
 TBRE $J=0, T=2$ & $22.1$ & $4.3$ & $37.1$ & $0.59$ & $0.12$ \\
\hline
 TBRE $J=2, T=0$ & $20.3$ & $4.2$ & $38.2$ & $0.53$ & $0.11$ \\
\hline
RTBRE $J=0, T=0$ & $5.0$ & $1.96$ & $12.3$ & $0.41$ & $0.16$ \\
\hline
 GOE & $0.015$ & $3.17\times 10^{-3}$ & $1.0$ & $1.5\times 10^{-2}$ & 
$3.1\times 10^{-3}$ \\
\hline
\end{tabular}
\end{center}
\end{table}

This indicates that the TBRE is not ergodic and thus
 unfolding with the ensemble averaged level density is inappropriate.
We therefore proceed to adjust a polynomial of degree 7 
 to each spectrum individually and
use this individual density for the unfolding procedure.
\begin{figure}
\begin{center}
\includegraphics[scale=0.6]{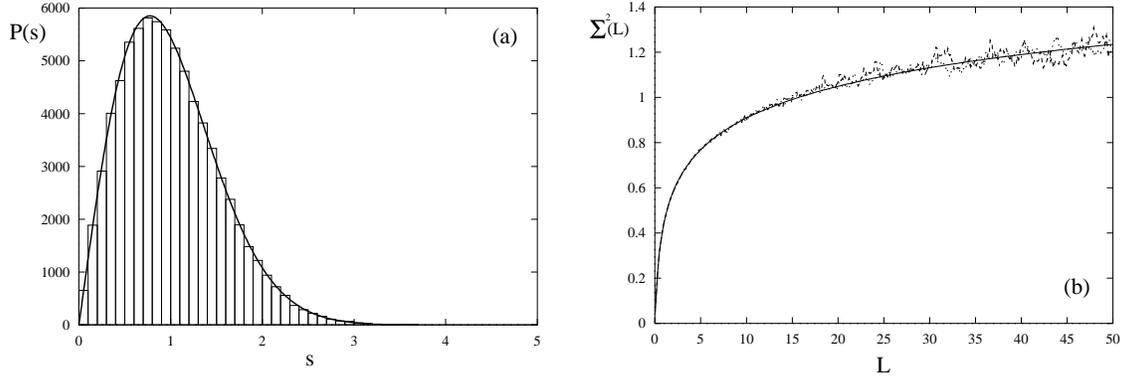}
\caption{(a) Nearest-neighbour spacing distribution of the TBRE
$(J=0, T=0)$ after each spectrum is unfolded individually.
The solid line corresponds to the GOE.
(b) Number variance $\Sigma^2(L)$ 
for the same case. 
The 
dotted and dashed curves correspond, respectively, to
spectral average and ensemble average after re-centering the spectra and
normalizing their widths. The full line is obtained for the GOE.}
\end{center}
\end{figure}

Figs 2a and 2b show that now $P(s)$ and $\Sigma^2(L)$ are in good agreement
with the GOE
prediction and Figs. 3a and 3b show a similar agreement
\begin{figure}
\begin{center}
\includegraphics[scale=0.6]{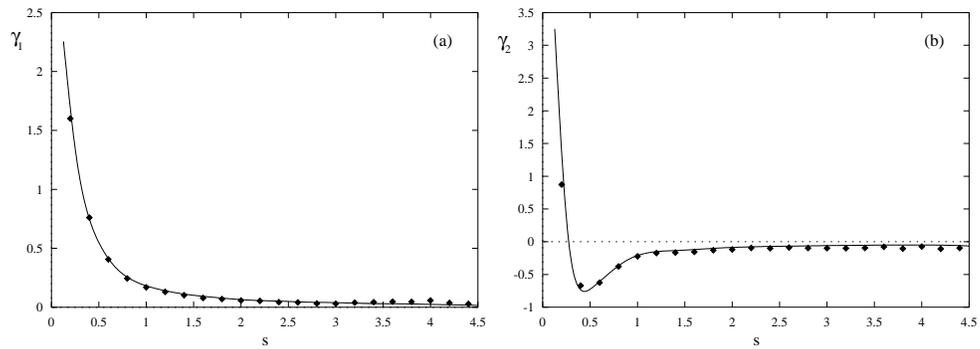}
\caption{(a) Skewness $\gamma_1$ and (b) excess $\gamma_2$
for the spectral average of the TBRE. The full line corresponds to the GOE.}
\end{center}
\end{figure}
for the skewness and excess. We should remark that in this case we could
increase the energy interval and decrease the degree of the polynomial
in order to get the high quality of the result for $\Sigma^2(L)$ up to $L=50$.
We discarded 85 states both at the upper and lower end of the spectra.
These numbers were chosen because omitting fewer states showed significant
edge effects while omitting more reduced the range of $L$ for which the
fluctuations
of $\Sigma^2(L)$ were acceptable.

We next wish to check whether the variations of the centers and widths of the
spectra were alone responsible  for the deviations
with respect to the GOE. Therefore we repeated the 
calculations with an ensemble unfolding after first re-centering all spectra
to the same value or, alternatively, normalizing their width to the same 
value.
The result for the number variance for both procedures is drawn in Fig. 1b.
Both curves for the number variance represent an improvement but
remain very far from GOE. Obviously we should apply both corrections, {\it 
i.e.} both re-center and dilate the spectra such as to correspond to 
uniform centers and widths before the ensemble unfolding. The 
result is such that on the scale of Fig. 1b it is not distinguishable from 
the GOE. The result is therefore plotted in Fig. 2b and is comparable to 
the one obtained by unfolding the individual spectra. 

On the average we discarded 105 states at the edges of the spectra and
found $m=15$ as the optimum degree for the polynomial in this analysis.
The
quality of the energy averaged $\Sigma^2(L)$ is only marginally  better.
Higher moments
play a minor role.
We obtained the same result for an ensemble of comparable size and quantum
numbers $J=0, T=2$ and $J=2, T=0$. 
Similar agreement was found for $P(s)$, as well as for skewness
and excess.

These results seem to indicate that the first and second moments are 
basically responsible for the non-ergodicity of the TBRE, but we have no 
theoretical argument to support this conjecture, and thus it might not 
always be true, or break down with higher exactitude.

The above procedure eliminates the large fluctuations in the average density
of each spectrum, but it is not totally equivalent to the procedure proposed
by French \cite{Fre73}. In the original paper the correction he calculated 
was given for the distribution  of widths of $n$-th neighbour spacings as a 
function of $L$. We translated the correction to one for
$\Sigma^2(L)$ (see equations (5.3) and (6.3) 
of reference \cite{Brodetal}) that reads as
\begin{equation}
\Sigma^2_s(L) = 
\frac{\Sigma^2_e(L) + \left [ \frac{1}{6} - L^2
\right ] \frac{\sigma^2}{D^2}
}{1 - \frac{\sigma^2}{D^2}}
\label{eqFre}
\end{equation}
where $\Sigma^2_e(L)$ and $\Sigma^2_s(L)$ denotes the ensemble and 
spectral (energy)
averaged number variance at distance $L$ and $\sigma^2$ denotes the ensemble
variance of the spectrally averaged mean level distance $D$.
Fig. 4  shows 
$\Sigma^2_s(L)$ as calculated 
from eq. (\ref{eqFre}) using the numerical data for $\Sigma^2_e(L)$.
\begin{figure}
\begin{center}
\includegraphics[scale=0.6]{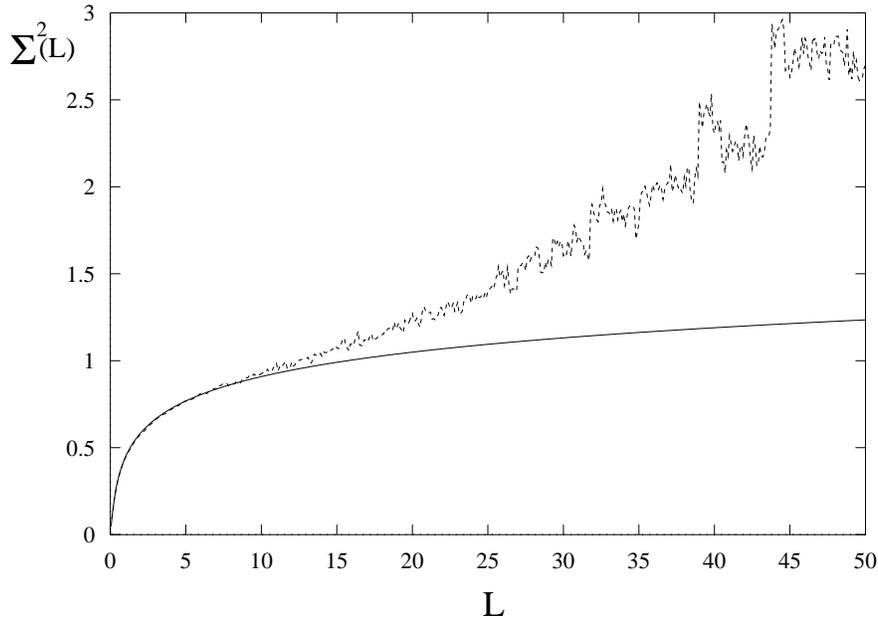}
\caption{Application of formula (\ref{eqFre}) as described in the text
(dashed line) in comparison
with the GOE (full line).}
\end{center}
\end{figure}
Note that this correction is applied after re-centering the 
spectra. This is not explicit in the original paper, but it is the only 
reasonable way
to interpret the basic argument given there. Indeed, application of the 
formula to non re-centred spectra gives unreasonable results.

For correlations up to about $L$=5 and certainly for the nearest-neighbour 
spacing distribution the correction works quite well; at larger distances 
discrepancies become large, as is to be expected if the aim is to simulate 
the adjustment of the width of a near Gaussian. The larger the spectra 
the larger the range on which we expect the correction to work well.
In earlier calculations with smaller ensembles the number variance was 
not 
considered at large distances, and thus the deviation could not be detected.

The spectral properties of the RTBRE at the center 
are also found to be of GOE type up to some energy
range which depends on the subspace defined by the quantum
numbers we consider and on the interaction strength.

\section{Properties at the edge of the spectra}

The fluctuations at the edge of the spectra are of interest 
because of the comparison with low-lying nuclear states \cite{BroCoFloMe},
and in the context of recent work on the angular momentum dependence 
of ground states \cite{Ber1,AFra,Ber2}. The quantity to consider is 
the nearest-neighbour spacing
distribution. The old results are fitted 
with a Brody distributions \cite{Brod73} 
with parameters around $0.5$. Recent 
studies have suggest the use of 
semi-Poisson distributions \cite{Shklo}, that might be relevant 
as we are interested in the edge of the spectra. These distributions 
have
been discussed in relation with triangles with irrational relations 
between angles \cite{Bogom}, where the coincidence is spotty. We shall keep this 
additional option in mind to expand on the old discussions.

The simplest way to proceed is to take all spacings between the first and
the second level  and normalize them to average 1
(for improved statistics we shall use the
corresponding spacings at the upper end of the spectrum also, without mentioning
it explicitly each time).
\begin{figure}
\begin{center}
\includegraphics[scale=0.6]{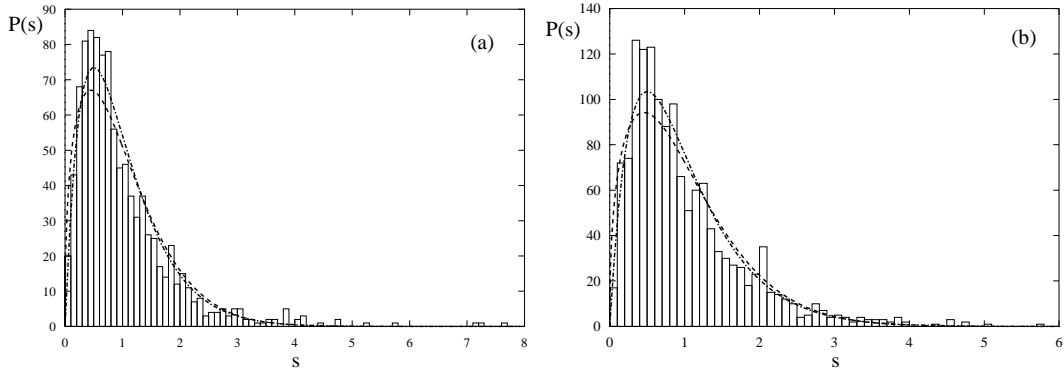}
\caption{ (a) Distribution of the spacings between ground and  first excited
states for the TBRE ($J=0,T=0$). The dashed-dotted line corresponds to
the semi-Poisson and the dashed line to the best Brody fit. 
(b) The same as in (a) for the quantum numbers ($J=2,T=0$)}
\end{center}
\end{figure}
In Fig. 5 we show the spacing distribution obtained in this way, and we find
amazingly large spacings up to 8 
in the case $J=0,T=0$ and up to 6 for 
the ensemble with quantum numbers $J=2,T=0$.
Note that the same procedure performed 
on a GOE of the same size leads to a spacing distribution very similar to 
that 
at the center of the spectrum. 
The semi-Poisson distribution does not yield a very convincing fit,
but is not worse than the usual Brody distribution, despite of the 
fact that the latter has one free parameter.

Normalizing the distances to $1$ might seem too na\"{\i}ve. 
Although the polynomial
unfolding of the integrated spectrum delivered good results for the 
analysis of the central region
it is not useful at the edges. Therefore we proceed to use a Gaussian
to unfold each spectrum individually adjusting its parameters.
This procedure has some merit; a Gaussian is the exact
result for the ensemble with infinite dimension, though the fact that we 
have already re-confirmed
the non-ergodicity of the TBRE casts some doubt on its application to 
individual spectra.
\begin{figure}
\begin{center}
\includegraphics[scale=0.6]{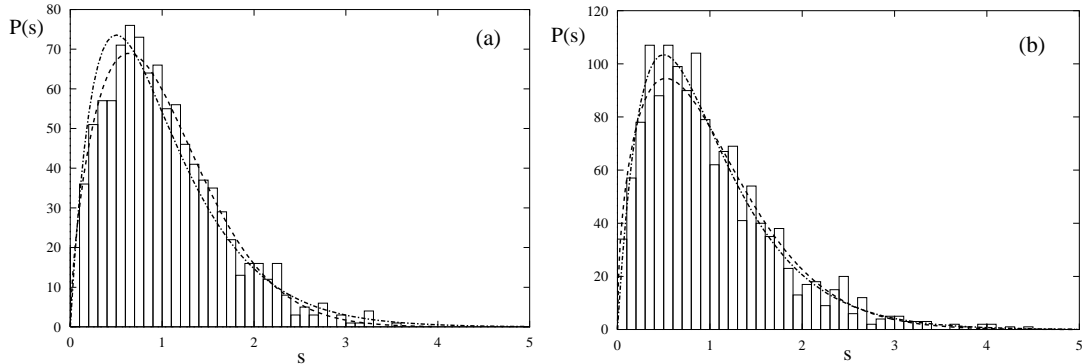}
\caption{ The same as in Fig. 5, but unfolding each spectrum individually
with a Gaussian.}
\end{center}
\end{figure}
In Fig. 6 we show the spacing distribution between the first 
and the second states for $J=0,T=0$ and $J=2,T=0$
unfolded in this way.
The extremely large spacings have disappeared though the result
still differs markedly from a GOE.
The confidence levels, discussed in \cite{Brod73}
for a Brody fit, 
rely heavily on binning the small intensities in the tail in
one large bin and are therefore somewhat arbitrary.
In particular, a few large spacings would not invalidate the fit.
The analysis of the ensemble 
for $J=0,T=2$ delivered qualitatively the same result and they agree 
substantially with the old results.
The semi-Poisson distribution drawn in the same picture gives a 
surprisingly good fit in Fig. 6b and is somewhat worse in Fig. 6a.
Note though that the discrepancies again are considerably smaller 
for the longer spectrum  and are not larger for the shorter one than for the 
Brody distribution with an adjusted parameter.

Yet we may ask if the Gaussian unfolding is really that meaningful. In this 
context 
two questions arise: First, is the procedure stable? And 
second, does 
it relate to the only experimental data we have? While the latter point 
will be discussed later, the former may be considered by
using a Gaussian multiplied by a polynomial for unfolding.
We may then ask if we find any plateau of the result as a function of the 
degree of the polynomial. We considered the Brody parameter as a function of 
the degree of the polynomial and we find that for the short spectrum a polynomial 
up to degree 2 and for the longer one a polynomial up to degree 4 can
be added without a significant change of the Brody parameter. 
This, combined with the theoretical expectation of a Gaussian in the limit of large
matrices, gives us some confidence to use the unfolding mentioned and should
encourage further analysis as to the significance of the agreement found
with a semi-Poisson distribution. 

We next look at the comparison with the nuclear physics results of
ref. \cite{BroCoFloMe}. Note
that here the normalization of the spacings is made as a function of 
mass number and angular momentum, but in no way is the spectral density 
further
up in the spectrum involved. We therefore conclude that only a comparison 
with the simply normalized spacings shown in Fig. 5 is legitimate. Yet 
these data have a very long tail that seems incompatible with the nuclear
data. As we discarded a more sophisticated unfolding, we have to fall back 
on an alternative option. Obviously the assumption of degenerate
single-particle levels is not fulfilled in the nuclear case. It therefore seems 
justified to consider realistic splitting of the
single-particle levels combined with a realistic interaction strength, i.e.
the RTBRE. In this case we must limit ourselves to the spacings at the 
bottom of the spectrum, as symmetry is destroyed by the single-particle 
energies.  Note that the results we now obtain are dependent on the shell 
splitting and 
by no means universal. Yet if we compare these results in Fig. 7
with the experimental ones we see less discrepancies. 
\begin{figure}
\begin{center}
\includegraphics[scale=0.6]{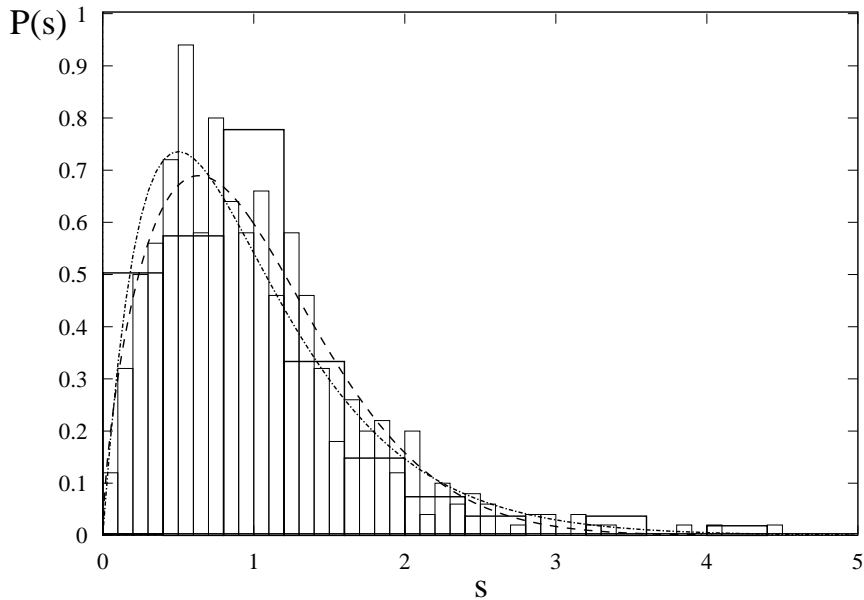}
\caption{ Distribution of the spacing between ground and first excited
states. The solid line histogram corresponds to the nuclear table
ensemble and the dashed line one to the RTBRE $(J=0, T=0)$.
The dashed-dotted line corresponds to the semi-Poisson 
and the dashed curve to the best Brody fit.}
\end{center}
\end{figure}
As the nuclear data have poor statistics the agreement
is not very meaningful, but at least the few large spacings are consistent and
the 
general shape seems correct. Fits with a Brody distribution fails in this case
and the agreement with a semi-Poisson distribution is similarly poor.
Comparison with the unfolded TBRE for the smaller sample, on the other hand
would give a better fit, but has no real foundation.

\section{Conclusions}

We may conclude that the phenomenology observed in earlier papers for the TBRE
is essentially correct when viewed with larger spectra. The
correction formula proposed in \cite{Fre73} will only correct short-range 
behaviour if we re-center the spectra previously.
We found that re-centering spectra and normalizing their widths would
remove the non-ergodicity at the  center of the spectrum to a large extent.
Note though, that this non-ergodicity is 
not really a problem at least if we consider the center of the spectra, 
because in practice we have to unfold each nuclear, atomic, resonance
cavity or whatever spectrum individually, anyway. 
At the center of the spectrum the fluctuation properties of the RTBRE
coincide with those of the TBRE on the range determined by the strength 
of the interaction.

As far as edge effects 
are concerned, the situation changes. 
Spectral properties with realistic splitting of single-particle levels differ 
significantly from those of the TBRE. The RTBRE seems to be in better agreement 
with the nuclear data, but no doubt a new compilation of such data is
desirable because the small sample
does not allow a detailed comparison.
The unfolding procedures at the edge of the spectrum are
quite sensitive, but we have found unfolding with a Gaussian to be quite satisfactory
both because of its theoretical background and because of its stability.
We certainly confirm earlier findings of large fluctuations, and with the Gaussian
unfolding we found very reasonable agreement with a semi-Poisson
distribution for the longest spectrum we have. 

We may conclude that future attention should be concentrated on two points:
First we should look out for possible effects of the non-ergodicity on wave function
properties, and second we should investigate the possible significance 
of the semi-Poisson distribution at the edge of spectra.\\

\noindent
{\small {\bf Acknowledgment:} We gratefully acknowledge valuable discussions 
with  F.~Leyvraz and I.~Rotter. 
This work is supported by the following grants:
CONACyT 32101-E, 32171-E,   
DGAPA (UNAM) IN-112998 and NSF 0070911
}

\end{document}